# Mixed reality technologies for people with dementia: Participatory evaluation methods

Shital Desai
Social and Technological Systems (SaTS) lab,
School of Arts Media Performance and Design,
York University, Toronto, Canada
desais@yorku.ca

Arlene Astell
KITE UHN Toronto Rehabilitation Institute
550 University Ave #12
Toronto, Canada
arlene.astell@uhn.ca

## ABSTRACT

Technologies can support people with early onset dementia (PwD) to aid them in Instrumental Activities of Daily Living (IADL). The integration of physical and virtual realities in Mixed reality technologies (MRTs) could provide scalable and deployable options in developing prompting systems for PwD. However, these emerging technologies should be evaluated and investigated for feasibility with PwD. Survey instruments such as SUS, SUPR-Q and ethnographic methods that are used for usability evaluation of websites and apps are used to evaluate and study MRTs. However, PwD who cannot provide written and verbal feedback are unable to participate in these studies. MRTs also present challenges due to different ways in which physical and virtual realities could be coupled. Experiences with physical, virtual and the couplings between the two are to be considered in evaluating MRTs. This paper presents methods that we have used in our labs – DATE and SaTS, to study the use of MRTs with PwD. These methods are used to understand the needs of PwD and other stake holders as well as to investigate experiences and interactions of PwD with these emerging technologies.

## 1. INTRODUCTION

Technologies can support people with early onset Dementia (PwD) to participate in Instrumental Activities of Daily Living (IADL) such as making a cup of tea, cooking and laundry. IADL is a list of activities related to independent living that health care professionals use to assess PwD for the level of impairment and their ability to care for themselves. PwD are unable to sequence tasks in an activity which makes it difficult for them to finish the task. Intelligent prompting systems can support PwD in completing IADL through prompts generated when PwD lose track of the activity (for example, A. Astell et al., 2009; Orpwood et al., 2008). Blended environments such as Mixed Reality Technologies (MRTs) could offer scalable and reconfigurable solutions that can be easily adopted and deployed.

MRTs consist of augmentations of physical and virtual elements and they come in various configurations [3]. Augmented reality and virtuality are two main categories of augmentations depending on whether physical is augmented with virtual (augmented reality) or virtual is augmented with physical (augmented virtuality). Use of MRTs as intelligent devices have been explored with Microsoft Kinect [4], augmented reality (AR) HoloLens [5] and projection based systems [6]. However, for MRTs to be used as prompts, these technologies need to be studied and evaluated with PwD. Understanding the experiences and interactions of PwD with MRTs is important for adoption and acceptance of these technologies by PwD. Our research on designing MRTs for PwD has thus focused on investigating experiences of PwD with MRTs through the concept of presence in blended environments [7] and identifying interaction modalities that work for PwD using perception action model [8].

Ethnographic methods such as participatory and codesign methods, observations, interviews, focus groups andsurveys can provide useful insights into the needs and experiences of people. Experience is evaluated through observations of users carrying out certain tasks with the technology. Standardised measures such as the System Usability Scale (SUS) for apps and SUPR-Q for websites involves users reflecting on their experiences with the technology with open ended and detailed questions about features [9], [10]. The same measures are also used in the design and evaluation of apps and websites for PwD [11].

Designing for experiences with MRTs revolves around creating an illusion of being in a certain place or environment when you are physically situated in another place [12]. So, attempts are made to make the digital world ubiquitous to the user. However, all realities in the design should be observable and detectable by PwD. They should be aware of the reality with which they are interacting for successful perception and action loops, thus contributing to positive experiences with the technology [7]. Creating illusions or the feeling of being somewhere else creates confusion rather than enhanced positive experiences in the context of PwD using MRTs as assistive technologies. Desai et al further emphasise that studying experiences with MRTs involves understanding people's experiences with physical and digital space as well as the correspondences or couplings between the two. Direct access to elements or objects in these spaces and the natural flow of actions on these elements is important. The challenge is to facilitate all of these while keeping the mediating technology ubiquitous to the user.

Ethnographic methods could present challenges in eliciting information from PwD and thus in evaluating technologies to be designed for them. Some PwD may be unable to provide verbal or written feedback in interviews and surveys. Studies such as [7], [8], [13] have successfully used observation methods to investigate experiences and interactions of PwD with MRTs. We are developing research methods in our labs: Social and Technological Systems (SaTS) lab and Dementia Ageing Technology and Engagement lab (DATE) – where the primary objective is to allow vulnerable populations or those who cannot provide verbal and written feedback due to their impairments, to have a say in the entire design and developmental process of technologies. We will discuss these methods and our experiences with these methods at the workshop – 'Evaluating User Experiences in Mixed Reality'.

## 2. Cocreating experiences using Tungsten

We have used TUNGSTEN<sup>TM</sup> (Tools for User Needs Gathering to Support Technology Engagement) (http://tungsten-training.com), a set of practical tools for researchers and technology developers to involve older adults as experts in the technology development, testing and implementation process, from conception of ideas to adoption of products (Astell et al., 2020). We have used these tools in half day and full day workshop settings to allow participants to share their experiences with technologies with all stakeholders [14], [15]. Older adults with dementia, their care givers, technology developers and health care professionals engaged in three TUNGSTEN co-creation activities (Figure 1): (i) Technology Interaction - activity designed to determine factors that influence older adults' impressions of new technologies from a 'mystery box' and that will enable them to persevere with trying to get them working and not abandon them, (ii) Show and Tell - activity designed to understand what makes people love or abandon technology that they have owned in the past or they own currently and (iii) Scavenger Hunt - is used to gather early feedback on a prototype, make it ready for market release and want to understand how users interact with products that are under development.

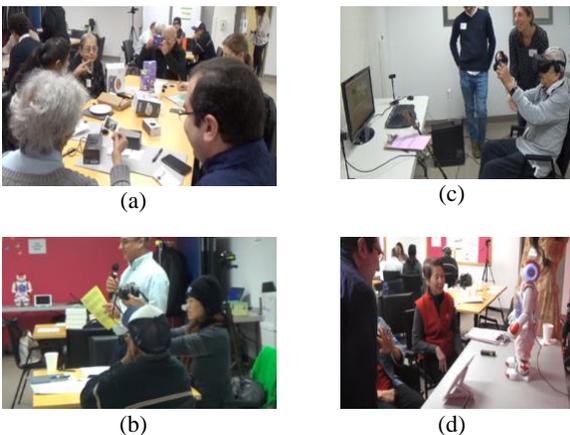

**Figure 1 People with early onset dementia participating in (a) Technology Interaction (b) Show and Tell (c) and (d) Scavenger Hunt**

## 3. Observation method

We used off the shelf MRTs – HoloLens and XBOX Kinect from Microsoft, Osmo from Tangible Play and ARkit from Apple (IphoneX) in our studies. Using off the shelf existing technologies is an effective way to understand technology needs of people and their perception action behaviour [7], [16]–[18]. We have used game play as a probe to elicit natural behaviour in the participants when they interact with MRTs. Games can also be easily integrated in the day programs of PwD. Play also acts as an ice breaker and makes participants feel more comfortable around emerging technologies such as MRTs. PwD played Tangram on Osmo, Young Conker on HoloLens and a game of bowling on XBOX Kinect and Stack AR IphoneX ARkit (Figure 2).

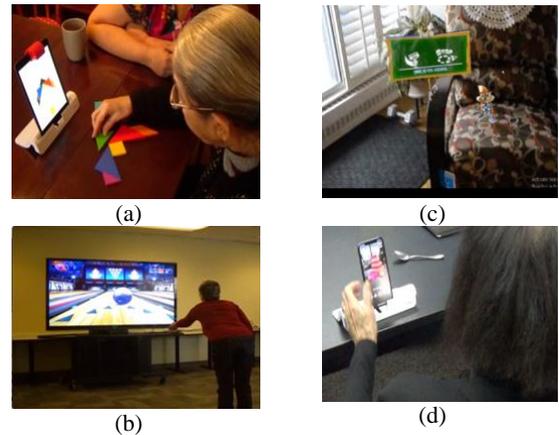

**Figure 2 (a) Tangram on Osmo (b) Bowling on Kinect XBOX (c) Young Conker on HoloLens (d) Stack AR on IphoneX**

Cognitive impairment of the participant is recorded using assessment tools such as MoCA before the game play sessions. The observations are video recorded for analysis in Noldus Observer XT, a software for analysis of behavioral data. It facilitates coding and description of participant behaviour over a period of observation time. The coding heuristics can either be determined deductively before the data collection, based on a theoretical framework or determined inductively during the analysis from the data. The coded data is then analysed either qualitatively using visualisations in Observer XT or quantitively using statistics or both. Figure 3 shows the coding environment in Observer XT, where data collected simultaneously from maximum four sources can be analysed at a given time.

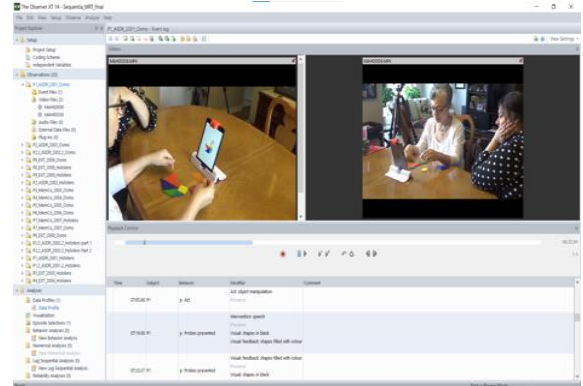

**Figure 3 Coding multiple video sources in Observer XT**

## 3.1 Triangulation with biological and egocentric data

In our studies with children [17], we have successfully used retrospective interviews [19] and concurrent and retrospective verbal protocols [20] in observational studies to reliably identify the behaviour codes in the data for thematic analysis. With PwD, some participants provided limited verbal protocols during the game play, but most did not provide verbal feedback. Thus, we are exploring use of additional data sources such as gaze data using eye tracking glasses, facial emotions using a face reader software and biological signals using EEG in addition to behavioural data. Figure 4 shows gaze information captured using eye tracking glasses.

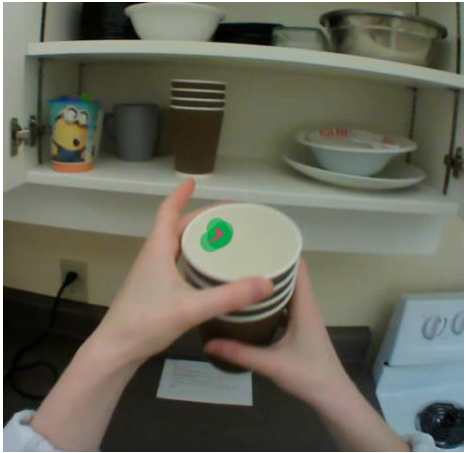

**Figure 4 Gaze information captured using eye tracking glasses in IADL: making cup of tea**

At any given time during participants' use of MRTs, data is captured from five sources: (1) video cameras capture behavioral data to determine actions and perceptions with the technology (2) eye tracking glasses capture gaze and pupil data to determine where participants are looking (3) FaceReader module from Observer XT indicates emotions of participants (4) EEG data provides quantitative information about neurological processes in the brain. (5) a task assessment tool created using Assessment of Motor and Process Skills (AMPS) [21] and the Perceive: Recall: Plan: Perform (PRPP) [22] is used to assess the execution of tasks in PwD with or without MRT support. Triangulation of all these data in Observer XT environment helps us to develop an exhaustive coding scheme for thematic analysis and also helps us to reliably code behaviours and interactions of participants with MRTs (Figure 5)

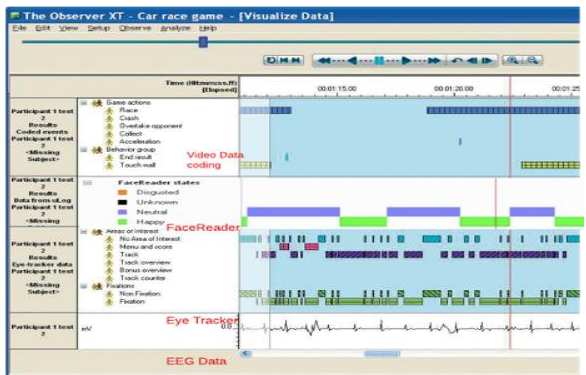

**Figure 5 Observer XT environment showing simultaneous visualisations of coding for data from video cameras, eye tracking glasses, FaceReader module and EEG**

## 4. Conclusion

Emerging technologies such as MRTs can support PwD in carrying out IADL. However, these technologies should be studied and evaluated with primary users and other stake holders. The impairments of PwD and the dual reality experienced in MRTs present challenges to the use of conventional methods in studying and evaluating MRTs with PwD. We have presented some of the methods that we use in DATE and SaTS lab to study MRTs with PwD. These methods are unique in the way that they can be adapted to the participant's abilities and impairments.

## 5. ACKNOWLEDGMENTS

The research described is funded by AGE-WELL. We are very thankful to all participants and the staff at Alzheimers Society of Durham, Memory and Company and Carefirst for their ongoing support in our studies. Thanks to Noldus for supporting our research with remote Observer licenses during the pandemic.